\documentclass[conference]{IEEEtran}
\IEEEoverridecommandlockouts
\usepackage{cite}
\usepackage{amsmath,amssymb,amsfonts}
\usepackage{algorithmic}
\usepackage{graphicx}
\usepackage{textcomp}
\usepackage{xcolor}
\def\BibTeX{{\rm B\kern-.05em{\sc i\kern-.025em b}\kern-.08em
    T\kern-.1667em\lower.7ex\hbox{E}\kern-.125emX}}

\usepackage{hyperref}

\usepackage[ruled, linesnumbered]{algorithm2e}
\usepackage{cleveref}
\SetKwInOut{Input}{input}
\SetKwInOut{Output}{output}
\SetKwInOut{Name}{name}
\SetKwProg{Function}{Function}{:}{end}
\SetKw{Break}{break}

\usepackage{enumerate}
\usepackage{enumitem}

\usepackage{tikz}
\usetikzlibrary{positioning,decorations.markings,chains,arrows,arrows.meta,shadows,fadings,shapes,backgrounds,snakes,matrix,patterns,plotmarks,trees,mindmap, calc, tikzmark, fit}

\usepackage{pgfplots}
\usepackage{transparent}

\usepackage[section]{placeins}

\usepackage{subfigure}
\pgfplotsset{compat=1.5}
\usepackage{pgfplotstable}
\usepackage{listofitems}

\newcolumntype{?}{!{\vrule width 1pt}}

\usepackage{multirow}

\pgfplotsset{
  discard if/.style 2 args={
      x filter/.append code={
          \readlist\mylist{#2}%
          \foreachitem\z\in\mylist[]{%
            \ifdim\thisrow{#1} pt=\z pt
              \def\pgfmathresult{inf}
            \fi
          }
        }
    },
  discard if not in/.style 2 args={
      x filter/.append code={
          \readlist\mylist{#2}%
          \providecommand{\foundit}{0}
          \renewcommand{\foundit}{0}
          \foreachitem\z\in\mylist[]{%
            \ifdim\thisrow{#1} pt=\z pt
              \renewcommand{\foundit}{1}
            \else
            \fi
          }
          \pgfmathparse{\foundit == 1? \pgfmathresult : nan}
        }
    },
}

\makeatletter
\newcommand{\gettikzxy}[3]{%
  \tikz@scan@one@point\pgfutil@firstofone#1\relax
  \edef#2{\the\pgf@x}%
  \edef#3{\the\pgf@y}%
}
\makeatother

\tikzset{rectstate/.style={
      draw=black,
      rectangle,
      align=center,
      minimum width=4cm
    }
}

\tikzset{emptyNode/.style={
      draw=none
    }
}

\tikzset{edge/.style={
      ->,
      draw,
      semithick,
      shorten >=1pt,
      >=stealth'
    }
}

\tikzset{
  between/.style args={#1 and #2}{
      at = ($(#1)!0.5!(#2)$)
    }
}

\tikzset{
  rowlabel/.style={
      inner sep=2pt,
      draw=none,
      font=\fontsize{8}{8}\ttfamily,,
      node contents={\the\numexpr-1+\pgfmatrixcurrentcolumn\relax},
      alias=m\the\numexpr-1+\pgfmatrixcurrentcolumn\relax
    }
}

\tikzset{
  columnlabel/.style={
      minimum size=0pt,
      draw=none,
      node contents={\the\numexpr-1+\pgfmatrixcurrentrow\relax},
      alias=c\the\numexpr-1+\pgfmatrixcurrentrow\relax
    }
}

\tikzset{
  arrayrow/.style={%
      matrix of nodes,
      row 1/.style={nodes = {draw, minimum size=8mm, anchor=center, text width=6mm, align=center}},
      column sep=-\pgflinewidth,
      row sep=-\pgflinewidth,
      nodes in empty cells,
      row 2/.style={nodes=rowlabel}}
}

\tikzset{
  arraycolumn/.style={%
      matrix of nodes,
      column 1/.style={nodes = {draw, minimum size=7mm, anchor=center}},
      column sep=-\pgflinewidth,
      row sep=-\pgflinewidth,
      nodes in empty cells,
      column 2/.style={nodes=columnlabel}}
}

\tikzset{
  treenode/.style={draw, circle, text width=6mm, align=center}
}

\pgfplotsset{
  legend image with text/.style={
      legend image code/.code={%
          \node[anchor=center] at (0.3cm,0cm) {#1};
        }
    },
}
\tikzset{every mark/.append style={scale=1.5}}

\usepackage{csvsimple}

\usepackage{ifthen}
\usepackage{xargs}
\usepackage{keyval}

\definecolor{color1}{rgb}{1,0,0}
\definecolor{color2}{rgb}{1,1,0}
\definecolor{color3}{rgb}{0,1,0}
\definecolor{color4}{rgb}{0,1,1}
\definecolor{color5}{rgb}{0,0,1}
\definecolor{color6}{rgb}{1,0,1}
\definecolor{color7}{rgb}{0,0,0}

\newcommand{\colorone}{red}
\newcommand{\colortwo}{yellow}
\newcommand{\colorthree}{green}

\newcommand{\colorfive}{blue}

\definecolor{graphcol1}{rgb}{.7,0,0}
\definecolor{graphcol2}{rgb}{.7,.7,0}
\definecolor{graphcol3}{rgb}{0,.7,0}
\definecolor{graphcol4}{rgb}{0,.7,.7}
\definecolor{graphcol5}{rgb}{0,0,.7}
\definecolor{graphcol6}{rgb}{.7,0,.7}
\definecolor{graphcol7}{rgb}{0,0,0}

\newcommand{\graphcolone}{red}
\newcommand{\graphcoltwo}{yellow}
\newcommand{\graphcolthree}{green}
\newcommand{\graphcolfour}{turquoise}
\newcommand{\graphcolfive}{blue}
\newcommand{\graphcolsix}{purple}

\newcommand{\etal}{et al.~}

\newcommand{\MM}{Memory Model}
\newcommand{\mms}{memory models}

\newcommand{\gmm}{heterogeneous memory model}

\newcommand{\GMM}{Heterogeneous Memory Model}

\newcommand{\cmm}{hierarchial memory model}

\newcommand{\CMM}{Hierarchial Memory Model}

\newcommand{\af}{access frequency}

\newcommand{\afs}{access frequencies}

\newcommand{\afm}{access frequency model}

\newcommand{\AFM}{Access Frequency Model}

\newcommand{\Fswap}{\textsc{Swap}}

\newcommand{\Fcopy}{\textsc{Copy}}
\newcommand{\Ftempnode}{\textsc{TempNode}}
\newcommand{\Fmaxindex}{\textsc{MaxIndex}}

\newcommand{\mergesort}{\emph{Merge Sort}}

\newcommand{\Smergesort}{MS}

\newcommand{\pathsort}{\emph{Path Reorder}}
\newcommand{\Hpathsort}{Path Reorder}
\newcommand{\Spathsort}{PR}

\newcommand{\shellsort}{\emph{Shell Sort}}
\newcommand{\cyclesort}{\emph{Cycle Sort}}
\newcommand{\Hcyclesort}{Cycle Sort}
\newcommand{\insertionsort}{\emph{Insertion Sort}}

\newcommand{\buffmergesort}{\emph{Buffer Merge Sort}}
\newcommand{\Hbuffmergesort}{Buffer Merge Sort}

\newcommand{\SNbuffmergesort}{NBMS}
\newcommand{\SMbuffmergesort}{MBMS}

\newcommand{\nativebuffmergesort}{\emph{Native Buffer Merge Sort}}

\newcommand{\mapbuffmergesort}{\emph{Map Buffer Merge Sort}}

\newcommand{\shellmergesort}{\emph{Shell Merge Sort}}
\newcommand{\Hshellmergesort}{Shell Merge Sort}

\newcommand{\SNshellmergesort}{NSMS}
\newcommand{\SMshellmergesort}{MSMS}

\newcommand{\nativeshellmergesort}{\emph{Native Shell Merge Sort}}
\newcommand{\mapshellmergesort}{\emph{Map Shell Merge Sort}}

\newcommand{\Hnativepathsort}{Native Path Reorder}
\newcommand{\nativepathsort}{\emph{Native Path Reorder}}
\newcommand{\mappathsort}{\emph{Map Path Reorder}}

\newcommand{\SNpathsort}{NPR}
\newcommand{\SMpathsort}{MPR}
\newcommand{\Fpathsort}{\textsc{PathReorder}}

\newcommand{\mapreorder}{\emph{Map Reorder}}
\newcommand{\Hmapreorder}{Map Reorder}
\newcommand{\Fmapreorder}{\textsc{MapReorder}}

\newcommand{\repone}{Location Representation}

\newcommand{\reptwo}{Target Representation}

\newcommand{\accthresh}{\emph{Access Threshold}}
\newcommand{\Haccthresh}{Access Threshold}

\newcommand{\ratiothresh}{\emph{Ratio Threshold}}
\newcommand{\Hratiothresh}{Ratio Threshold}

\newcommand{\btree}{B$^+$-Tree}
\newcommand{\btrees}{B$^+$-Trees}

\newcommand{\bstree}{Binary Search Tree}
\newcommand{\bstrees}{Binary Search Trees}

\newcommand{\avltree}{AVL Tree}
\newcommand{\avltrees}{AVL Trees}

\newcommand{\octree}{Octree}
\newcommand{\octrees}{Octrees}

\newcommand{\fullmsp}{MSP430FR5994}
\newcommand{\msp}{MSP}
\newcommand{\fullepyc}{AMD EPYC 7742}
\newcommand{\epyc}{EPYC}

\begin{document}

\title{Realizing Hardware-Optimized General Tree-Based Data Structures for Heterogeneous System Classes}
\author{\IEEEauthorblockN{1\textsuperscript{st} Daniel Biebert}
\IEEEauthorblockA{\textit{TU Dortmund University} \\
Dortmund, Germany \\
\href{mailto:daniel.biebert@tu-dortmund.de}{daniel.biebert@tu-dortmund.de}}
\and
\IEEEauthorblockN{2\textsuperscript{nd} Christian Hakert}
\IEEEauthorblockA{\textit{TU Dortmund University} \\
Dortmund, Germany \\
\href{mailto:christian.hakert@tu-dortmund.de}{christian.hakert@tu-dortmund.de}}
\and
\IEEEauthorblockN{3\textsuperscript{rd} Jian-Jia Chen}
\IEEEauthorblockA{\textit{TU Dortmund University} \\
Dortmund, Germany \\
\href{mailto:jian-jia.chen@cs.tu-dortmund.de}{jian-jia.chen@cs.tu-dortmund.de}}
}

\maketitle

\thispagestyle{plain}
\pagestyle{plain}

\begin{abstract}
Tree-based data structures are ubiquitous across applications. Therefore, a multitude of different tree implementations exist. However, while these implementations are diverse, they share a tree structure as the underlying data structure. As such, the access patterns inside these trees are very similar, following a path from the root of the tree towards a leaf node. Similarly, many distinct types of memory exist. These types of memory all have different characteristics. Some of these have an impact on the overall system performance. While the concrete types of memory are varied, their characteristics can often be abstracted to have a similar effect on the performance. We show how the characteristics of different types of memories can be used to improve the performance of tree-based data structures. By reordering the nodes of a tree inside memory, the characteristics of memory can be exploited to optimize the performance. To this end, this paper presents different strategies for reordering nodes inside memory as well as efficient algorithms for realizing these strategies. It additionally provides strategies to decide when such a reordering operation should be triggered during operation. Further, this paper conducts experiments showing the performance impact of the proposed strategies. The experiments show that the strategies can improve the performance of trees by up to 95\% as offline optimization and 75\% as online optimization.
\end{abstract}


\section{Introduction}
\label{cpt:intro}

Tree-based data structures are ubiquitous across applications and fields.
Due to their intuitiveness and inherent capability to divide data naturally, they are widely used. From large indexing structures designed to efficiently store and retrieve billions of elements to machine learning models used for inference, many applications use trees as their underlying data structure \cite{DBLP:conf/damon/KuhnBHCT23, DBLP:books/wa/BreimanFOS84}.

Despite the large variety of trees, the tree implementations usually follow similar access patterns of the nodes in the tree. These similarities open a design space to optimize different types of tree implementations with the same strategies. One such strategy is exploiting knowledge about the memory in the system, to optimize tree operations towards their respective characteristics. This paper proposes strategies for the optimization of tree-based data structures using the underlying memory.

Research towards optimizing the performance of trees often concentrates on one type of tree implementation \cite{DBLP:conf/sigmetrics/HankinsP03, DBLP:conf/ifipTCS/SaikkonenS08, DBLP:conf/icde/LeisK013,paralleltree}. However, most tree-based data structure implementations have characteristics in common.
Their shared node structure means the way data is accessed usually follows a pattern of moving from the root of the tree in the direction of a leaf. Their similarity can be exploited to apply the same methods of optimization to most types of tree-based data structures.

One way specific tree implementations have been optimized is by manipulating the way the nodes of a tree are stored in system memory \cite{DBLP:journals/tecs/ChenSHBLLMC22, DBLP:conf/damon/KuhnBHCT23}. The way system memory is accessed has a significant impact on the performance of a program. Not only does the access to system memory influence the execution time, but it also affects other types of performance, such as energy efficiency. Trees are no exception and can be optimized for a concrete memory technology.

Similar to trees, different memory technologies share characteristics which influence performance \cite{memorytech}. These characteristics can be abstracted to influence an optimization goal similarly.
Even though the specific characteristic influences different metrics, the same type of strategies can be used when optimizing the specific metric.

One very common characteristic of systems which opens an opportunity for optimization is some form of memory hierarchy in the form of cache \cite{DBLP:books/daglib/0072458}. By manipulating the data in memory, the CPU cache can be used more effectively. This improves performance when repeatedly accessing data.

The objective of this paper is to optimize the performance of tree-based data structures by manipulating the order of the nodes in memory. Furthermore, this optimization should be as general as possible, to be applicable to as many distinct types of tree implementations as possible. Both the tree implementation and the resulting tree instance should remain unchanged and optimization is only applied ''on top'' of the already existing implementation. Similarly, the optimization should be independent of the memory technology enabling the use of the same optimization for different types of memory. The result should therefore be tree implementation and memory technology independent while optimizing as many scenarios as possible.

\section{Related Work}
\label{cpt:related}

The process of optimizing the placement of data in systems with heterogeneous memory has been heavily researched in the literature. Avissar \etal showed that optimal static allocation improves the runtime for embedded systems with static random access memory (SRAM) and dynamic random access memory (DRAM) \cite{DBLP:conf/cases/AvissarBS01}. Udayakumaran \etal expanded the approach to dynamically copy frequently accessed data into SRAM to increase speed \cite{DBLP:conf/cases/UdayakumaranB03}. A software alternative to hardware cache was developed by Hallnor \etal \cite{DBLP:conf/isca/HallnorR00}. Many software-managed approaches to dynamically allocate or move data between different types of memory (usually SRAM and another type of memory) have been proposed \cite{DBLP:conf/cases/AngioliniMFBO04, 1382603, DBLP:conf/cases/NguyenDB05,DBLP:journals/jec/DominguezUB05}. Verma \etal integrated an algorithm into the compiler, which analyzes an application for data access and places parts of an array into the Scratch Pad Memory (SPM) for improved runtime and energy efficiency \cite{DBLP:conf/aspdac/VermaSM03}. Cho \etal used a similar approach, analyzing irregular access patterns in arrays to optimize the placement of array contents into SPM \cite{DBLP:conf/cases/ChoIDYP07}.

Utilizing CPU cache and improving both cache locality and prefetching has been studied extensively. A compiler-based prefetching approach working on recursive data structures was proposed by Chi-Keung \etal \cite{DBLP:conf/asplos/LukM96}. A similar approach for optimizing prefetching on linked data structures was introduced by Roth \etal \cite{DBLP:conf/isca/RothS99}. Karlsson \etal show a prefetching technique which works on linked data structures with irregular access patterns \cite{DBLP:conf/hpca/KarlssonDS00}. Methods of reordering data in memory to improve cache locality and increase performance have been studied extensively \cite{DBLP:conf/ipps/Al-FuraihR98, DBLP:conf/pldi/DingK99, DBLP:journals/ijpp/Mellor-CrummeyWK01, DBLP:conf/IEEEpact/MitchellCF99, DBLP:conf/asplos/CalderKJA98}. Chilimbi \etal focused specifically on improving the cache locality of data structures connected through pointers \cite{DBLP:conf/pldi/ChilimbiHL99}.

Optimizing trees in the context of different memory characteristics has been studied for individual types of trees. \btrees~have been studied and variations proposed which perform better in the context of prefetching and cache locality \cite{DBLP:conf/sigmetrics/HankinsP03, DBLP:journals/jcst/LuanDW09, DBLP:conf/damon/KuhnBHCT23}. Binary trees, such as \avltrees~and Red-Black Trees were optimized by storing them in a cache-sensitive layout in memory by Saikkonen \etal \cite{DBLP:conf/ifipTCS/SaikkonenS08}.

Using the distribution of frequently accessed nodes to drive a placement policy for improved caching and prefetching has been realized using random forests and \btrees. Chen \etal showed that placing frequently accessed paths inside trees sequentially into memory can speed up the performance of inference in random forests \cite{DBLP:journals/tecs/ChenSHBLLMC22}. K\"{u}hn \etal applied the same concept as offline optimization to \btrees, achieving a speedup in the access times of read operations \cite{DBLP:conf/damon/KuhnBHCT23}.

\section{\MM}
\label{cpt:memory}

A multitude of different memory technologies are in use today \cite{memorytech}. Every technology exhibits different characteristics, that influence performance.
If used thoughtfully, they can be used to improve the performance of the system \cite{DBLP:conf/cases/AvissarBS01}. However, if used without care, the same characteristics might result in worse performance. Careful consideration of the underlying memory technology is therefore important.

Even though every technology has different characteristics, many of them can be abstracted to have a similar effect on performance.
This section defines the abstract \mms, which generalize the characteristics of different technologies in terms of system performance.
Two types of \mms~are distinguished. First, a general view, where a system is equipped with at least two types of distinct memory technologies is examined. Second, a more concrete case where the system utilizes some form of caching is explored.

Performance can be defined by different metrics depending on the context.
While performance is often understood as the execution time of a program, this definition is too narrow for the purposes of this paper. The performance of a system can also be concerned with other metrics, such as energy efficiency. In fact, depending on the circumstances, the performance of a system might mean any metric, which is supposed to be optimized.
Therefore, whenever we talk about optimizing the performance of a program, we refer to optimizing one or multiple metrics, the nature of which is not relevant to the context of this paper.

\subsubsection{\GMM}
\label{sec:memory-general}

From an abstract general view, any memory technology will have some characteristics which influence performance \cite{sousa2005non}.
The \gmm~assumes that the system is equipped with at least two types of memory technologies, with characteristics that influence the performance differently. Furthermore, the types of memory should be sortable by their benefit towards the performance goal. It is therefore known which type of memory is expected to improve performance relative to the optimization goal when used most often.

\subsubsection{\CMM}
\label{sec:memory-cache}

Many CPUs include some form of cache to benefit access latency of frequently used contents in memory \cite{DBLP:books/daglib/0072458}. The exact implementation varies in different CPUs and architectures and is often not known. However, the exact implementation is not relevant for the purposes of this paper.

CPU cache is used to reduce access latency when accessing frequently accessed contents in memory \cite{DBLP:books/daglib/0072458}. The cache has lower access latency than the main memory. By keeping frequently accessed contents in the CPU cache, repeated access to the same content reduces the access latency.

In addition to simply caching the contents being accessed, some form of prefetching is usually implemented \cite{DBLP:books/daglib/0072458, DBLP:conf/isca/Jouppi90}. The task of prefetching is to not only load the content necessary, but also content which is expected to be needed in the future. Again, the exact method of prefetching is different depending on the CPU. However, a common way to improve the prefetching effectiveness is to exploit linear access to content in memory and reduce the distance of content inside memory. Placing content that is accessed sequentially linearly inside memory results in a recognizable access pattern and helps the content to be prefetched more easily. Additionally, reducing the distance of data accessed shortly after each other increases the likelihood of the data being prefetched together.

\section{Tree-Based Data Structures}
\label{cpt:tree-implementation}

Tree-based implementations are widely used across most fields. Being an intuitive way of structuring and dividing data, trees are used in fields ranging from databases \cite{DBLP:conf/damon/KuhnBHCT23, DBLP:conf/icde/LeisK013,DBLP:journals/jcst/LuanDW09} to machine learning \cite{DBLP:books/wa/BreimanFOS84}. However, while trees are versatile and plenty of implementations exist, most follow a similar pattern of data access.

\subsection{Tree Data Access Pattern}
\label{sec:tree-access-pattern}

Tree implementations will have some tree operations. Tree operations are defined as any atomic operation, accessing or changing the structure and data of the tree. They are atomic in the sense, that interrupting the process at any earlier point than completion would result in no progress being made. An example of a tree operation is finding and returning a value from a tree.

Due to the inherent structure of the tree, these tree operations follow a similar pattern.
An operation starts by accessing the root of the tree. 
It then continues to do some further operations on the root node, resulting in either a break condition being met, or determining one of the children as the next node to examine.
This pattern is repeated until the break condition is met and the operation ends. The break condition is often reaching a leaf node. It can however also be different, resulting in the operation ending in an inner node.
Actual implementations can be more complex than the outlined pattern.
However, due to a node only ever first being accessible by its parent (when starting from the root of the tree), the access pattern always follows paths, which move into the direction of the leaves.

\subsection{\AFM}
\label{sec:tree-access-frequency}

The \afm~of a tree describes the distribution of accesses.
The \af~of a node is chosen to be the number of access to this node.
After the tree was in use for a while and the \afs~have been counted, the \afm~can be used to determine the most accessed nodes. These are the ones with the highest \afs. Assuming that the future access pattern does not change significantly after this point, these nodes can be assumed to be accessed most often in additional tree operations.

\section{Node Reordering}
\label{cpt:reorder}

The core objective of this paper is to reorder tree nodes inside memory without changing the logical tree structure to utilize the benefits of the specific memory. To this end, two \mms~have been outlined.
Different reordering strategies are possible for the two models.
This section discusses different reordering strategies of nodes in memory.
In the following, whenever mentioning a reordering strategy, the overall intended order is meant. A reordering algorithm is one specific implementation, realizing the intended order of a reordering strategy.

The new order should be realized as performant as possible. Whenever a reordering operation is started some performance is lost in the form of the overhead caused by the reordering operation. To reduce this effect as much as possible, the reordering algorithms should be optimized towards performance, to cause the least possible overhead.
In most reordering algorithms copying a node from one memory location into another will dominate the performance impact. Depending on the specific implementation of a node, the impact varies. However, it always includes copying multiple values from one location to another, as well as updating the pointer in its parent. This is costly as almost all data of the node needs to be accessed and moved.
As copying is the most performance-sensitive operation, minimizing the number of copy operations is at the core of optimizing the reordering algorithms.
The algorithms should furthermore reorder in place to use as little additional memory as possible.

\subsection{Towards Reordering}
\label{sec:reorder-towards}

Before explaining the strategies for reordering the nodes, some general modifications to the way nodes are stored have to be made. To ensure more straightforward reordering, the nodes should be stored continuously. However, in the case of the \gmm~this is not possible, as some nodes are stored in another type of memory than others. To still have a storage scheme, which is similar to continuous, the different areas in the types of memory are combined into an abstract array. The abstract array is not physically stored anywhere, but maps the different physical locations in the types of memory into a logical array location. For this, the areas in the different memories are placed after one another inside the abstract array.

Any tree node can therefore be described by three distinct locations. First, the logical tree location. Second, the logical array location indicated by its index. And third, the physical location, which is its pointer.
Knowing the length and location of all arrays in all types of memory, the logical location can always be resolved to its physical location. Similarly, a physical location can also always be resolved to its logical location. Therefore, both sufficiently describe the location of a node in memory. For simplicity, the logical array location is used to describe the location of the node inside memory in the following.

The abstract array is filled successively with nodes whenever a new node is created. The first type of memory is therefore filled first. Once it is filled, the second type is used until it is filled. This is continued for all types of memory. Therefore, the order the types of memory are put in, in the abstract array, should be chosen carefully. By ordering the types of memory by their benefit towards the optimization goal, the most beneficial type of memory is filled first.

During reordering, the physical location of two nodes often needs to be swapped. This can be done using three copy operations. Every copy operation needs to ensure to update the pointer in the parent node to the new location.

\subsection{Reordering in the \GMM}
\label{sec:reorder-gmm}

\begin{figure}
    \centering
    \includegraphics{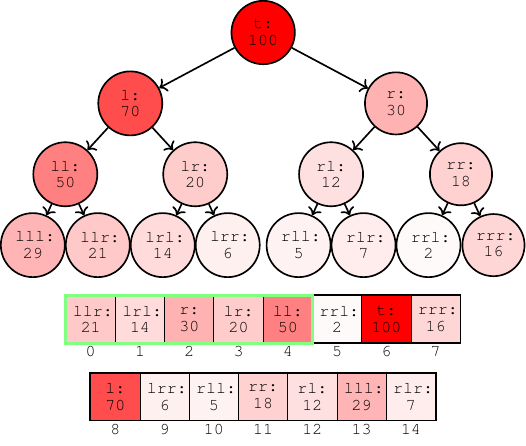}
    \caption{Naturally Grown Tree with Array Locations}
    \label{fig:reorder-gmm-example-before}
\end{figure}

In the \gmm, the arrays are already sorted by their impact on the performance goal. This ensures that the more beneficial memory types are filled before any less beneficial memory types. However, the order of the nodes inside the abstract array can still be arbitrary depending on the order the nodes are created and the tree implementation used. An example tree with the locations inside the array is given in \Cref{fig:reorder-gmm-example-before}.
The nodes are colored by their access frequency. A higher \af~results in a darker \colorone. Both the tree representation and the memory representation in the abstract array are shown.

As can be seen due to the way the tree grew ''naturally'', the nodes are spread across the abstract array. The part of the abstract array marked in \colorthree~is placed in a more beneficial type of memory. This memory should be accessed as often as possible. The most accessed nodes should therefore be placed in this type of memory. This can be done by sorting the nodes inside the abstract array according to their access frequency.
This reordering strategy, referred to as \mergesort~(\Smergesort), is chosen for the \gmm.

\begin{figure}[h]
    \centering
    \includegraphics{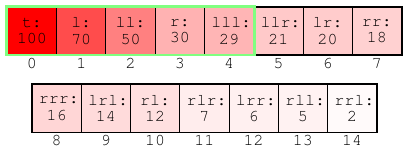}
    \caption{Sorted Array Locations}
    \label{fig:reorder-gmm-example-after}
\end{figure}

The resulting order for the example tree from \Cref{fig:reorder-gmm-example-before} can be seen in \Cref{fig:reorder-gmm-example-after}. The tree structure is not changed, as the nodes should only be reordered in memory. However, the most accessed nodes are now placed in the beneficial type of memory.

To fulfill the objective of reordering the nodes as efficiently as possible, \mergesort~is picked as a sorting algorithm. However, in the standard version of \mergesort, a copy of the contents to sort is made. This conflicts with the objective of reducing memory usage as much as possible. Therefore, in place variants of \mergesort~are chosen.

Multiple candidates exist, which sort the array in place, using a similar method to a standard \mergesort~implementation. Two variants have been picked with different competing characteristics. These two variants are explained in the following.

\subsubsection{Merge Sort - \Hbuffmergesort}
\label{subsubsec:reorder-gmm-merge-buffer}

The first version uses an internal buffer \cite{kronrod1969optimal}. It designates parts of the array as an internal buffer, in which the order of the elements can be arbitrary. This buffer can then be used to merge the rest two sorted areas in the rest of the array into. Afterwards, parts of the buffer can be sorted, decreasing the length of the buffer. Once most of the array is sorted and the buffer is sufficiently small the last elements of the buffer are ordered into the array at the correct location using another sorting algorithm such as \insertionsort.
This method was first proposed by Kronrod \cite{kronrod1969optimal}. Many adaptations of the idea exist \cite{DBLP:journals/ipl/Chen06,DBLP:journals/tcs/GeffertKP00,DBLP:journals/cacm/HuangL88,DBLP:journals/cj/HuangL92,DBLP:journals/ipl/MannilaU84,baeldung}. The version used in this thesis is taken from \cite{baeldung}.
For an array of size $n$, \buffmergesort~uses $\mathcal{O}(n * \log(n))$ number of copy operations in the worst case.

\subsubsection{Merge Sort - \Hshellmergesort}

As a second variant of in place \mergesort, \shellmergesort~is introduced. \shellmergesort~modifies the merge function of a standard \mergesort~implementation to an adaptation of \shellsort~\cite{DBLP:books/garland/Pratt72}. The idea of \shellsort~is to use \insertionsort, however rather than comparing and swapping the immediate next element, far apart elements are considered first. As the two parts are already sorted, \shellsort~can be used to reduce the worst-case number of swap operations. The concrete version used in this paper is taken from \cite{geeksforgeeks}.
\shellmergesort~results in a worst-case number of copy operations of $\mathcal{O}(\log(n) * n * \log(n))$.

\subsubsection{Merge Sort - Discussion}
\label{subsec:reorder-gmm-merge-discussion}

This section introduces two competing variants of \mergesort.
While \buffmergesort~has a worst-case number of copy operations of $\mathcal{O}(n * \log(n))$, \shellmergesort~has a worst-case number of copy operations of $\mathcal{O}(\log(n) * n * \log(n))$. Taking only this into consideration, \buffmergesort~seems to be the better choice, as it has the lower worst-case number of copy operations and better fulfills the objective of sorting as performant as possible. However, the worst-case number of copy operations is not necessarily the only relevant metric.
In many situations, parts of the node array might already be sorted. Especially if the tree had already been sorted previously and only parts of the \afm~changed, most of the array might still be correctly sorted, only needing a few swaps to be fully sorted again.
This situation does not favor \buffmergesort, because using the buffer always results in many swap operations being executed, even in the best case. The algorithm therefore does not benefit from an already partially sorted array.

\shellmergesort~on the other hand compares the element values during merging and only executes a swap operation if necessary. Parts of the array being sorted can therefore reduce the number of swap operations significantly. \shellmergesort~might therefore result in fewer copy operations in realistic use cases compared to \buffmergesort.

\subsection{Reordering in the \CMM}
\label{sec:reorder-cmm}

The \cmm~opens a wider design space for how to order nodes inside memory. As all memory is assumed to have the same properties in the \cmm, simply sorting into the ''better'' and ''worse'' memory cannot be used as a strategy. Rather, the distance of nodes, which influences spatial locality, should be regarded. The explained order is based on the order explored for random forests in \cite{DBLP:journals/tecs/ChenSHBLLMC22} and \btrees~in \cite{DBLP:conf/damon/KuhnBHCT23}.

An intuitive strategy of increasing spatial locality (i.e. decreasing the distance of frequently accessed nodes) is reusing \mergesort~as explained in the previous section. This ensures that the most accessed nodes are close together, which helps the nodes to be cached and prefetched more effectively for successive accesses to the same nodes.
However, this is a naive view of the accesses to tree nodes and fails to consider the concrete access patterns in trees. Nodes are always accessed in paths from the root node to one of the leaves. Nodes in different subtrees are therefore usually not accessed in the same tree operation.

\begin{figure}
    \centering
    \includegraphics{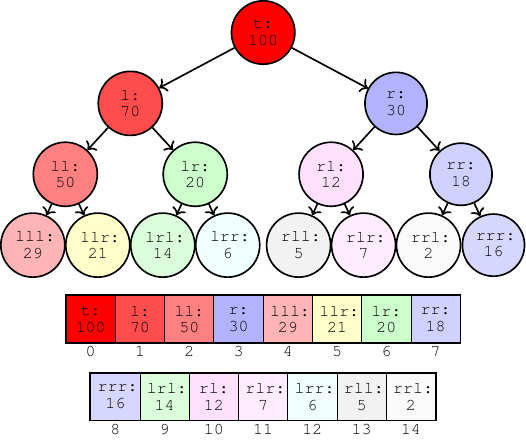}
    \caption{Sorted Tree with Array Locations - Colored Paths}
    \label{fig:reorder-cmm-example-mergesort-paths}
\end{figure}

To show this concept more clearly, \Cref{fig:reorder-cmm-example-mergesort-paths} shows the same tree and order as \Cref{fig:reorder-gmm-example-after} (i.e. sorted by their \af), however, with each most accessed path being colored in a different color. The saturation still shows the \af~of each node. For example, the most accessed path from \texttt{t} to \texttt{lll} is colored in \colorone~and the second most accessed path (without the nodes on the more frequently accessed path) from \texttt{r} to \texttt{rrr} is colored in \colorfive.

In this example \texttt{r} is placed directly next to \texttt{ll} because it is the next most accessed node. However, in an actual tree operation, \texttt{r} would not be accessed after \texttt{ll}. Either \texttt{lll} or \texttt{llr} would be accessed with \texttt{lll} having a higher \af~and therefore being more likely to be accessed next.

A better order considers this pattern. As nodes are always accessed in paths, the nodes should be placed sequentially inside memory. This results in a linear access pattern in memory, which helps the nodes to be prefetched more effectively. However, every node is only saved once in the array, therefore not all complete paths from the root to a leaf can be placed in memory sequentially. The \af~of nodes needs to be considered, placing the frequently accessed paths first and continuing with all not yet placed nodes.

\begin{figure}
    \centering
    \includegraphics{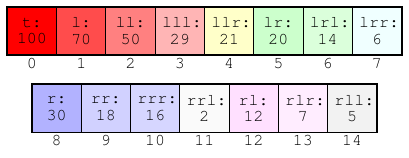}
    \caption{Reordered Tree using \Hpathsort~with Array Locations}
    \label{fig:reorder-cmm-example-after}
\end{figure}

An example of the resulting order is shown in \Cref{fig:reorder-cmm-example-after}. Here, the most frequently accessed path, marked in \colorone, is first placed sequentially in the array.
Afterwards, the next path to be placed in memory needs to be chosen. One option is to choose the second most frequently accessed path, marked in \colorfive. If this path is taken by the tree operation, its physical proximity to the already placed path does not benefit its execution. When the path colored in \colorfive~is taken, the other nodes in the \colorone~path are not accessed, the only exception being the root. However, the root is placed at the beginning of the array, far away from where the \colorfive~path would be placed. A better candidate to place after the \colorone~path is therefore the path colored in \colortwo. Right before the \colortwo~path is taken, most of the \colorone~path was accessed. Its proximity to the nodes on the \colorone~path therefore benefits spatial locality in an actual tree operation. Similarly, the next path to be placed is the path colored in \colorthree, as this again benefits spatial locality to the path colored in \colorone. The overall order therefore backtracks up the tree and places the most frequently accessed path which is not placed yet.

As mentioned, this strategy of reordering trees by placing the most frequently accessed paths sequentially in memory is not novel and was explored for random forests and \btrees~in literature \cite{DBLP:journals/tecs/ChenSHBLLMC22,DBLP:conf/damon/KuhnBHCT23}. However, the efficiency of the reordering algorithm was not considered, but rather only the effects on the performance of tree operations as offline optimization. Therefore, this paper uses the reordering strategy referenced as \pathsort~(\Spathsort) and introduces an efficient algorithm realizing the order. The following explains the algorithm, as well as analyzes its number of copy operations and memory requirements.

\subsubsection{\Hpathsort}
\label{subsec:reorder-path-method}

\begin{algorithm}
    \Function{$\Fpathsort(nodes, node, loc)$}{
    $sn = nodes[loc]$\;   \label{line:path-sort-find-swap}
    $\Fswap(node, sn)$\;   \label{line:path-sort-swap}
    $loc = (loc + 1) \% \Fmaxindex()$\;
    $ca = sn.\textsc{Children}()$\; \label{line:path-sort-children}
    $\textsc{Sort}(ca)$\;         \label{line:path-sort-merge}
    \ForEach{$c \textbf{ \emph{in} } ca$}{
        $\Fpathsort(nodes, ca, loc)$\;
    }
    }
    \caption{\Hpathsort} \label{alg:path-sort}
\end{algorithm}

The idea of \pathsort~is to use the knowledge of the structure of the tree to place frequently accessed paths sequentially in memory.
The algorithm is shown in \Cref{alg:path-sort}. The strategy of the \pathsort~algorithm is to place the current node at the next not yet reordered location and recursively call the function in the order of the most accessed children. It takes the node array $nodes$, the current node $node$ and the current location $loc$ as arguments. Notably, $loc$ is a static variable that references the same value across all function calls, so it reflects changes at another point in the recursion. Initially, the current node is initialized to be the root and the current location is initialized to be 0.

The algorithm starts by finding the node at the location $node$ is supposed to be placed at (\Cref{line:path-sort-find-swap}). This node is called $sn$. Afterwards, $node$ and $sn$ are swapped (\Cref{line:path-sort-swap}). Then the location counter is increased by one, making sure to handle an overflow of the array by starting from the beginning again. When initializing $loc$ to 0 this is not necessary. However, in later discussed situations $loc$ is not initialized as 0 and in these situations, the overflow needs to be handled.

Having placed the current node (now at $sn$) at the intended location, an array with its children is loaded in \Cref{line:path-sort-children}.
This sorted array is used in the next step to recursively call each child in the order of their access frequency. This ensures that the next most accessed path when moving up the tree is always placed next in the array. The recursion ends naturally in a leaf, as a leaf does not have any children and no recursive call is made.

In \Cref{alg:path-sort}, every node is visited exactly once, when it is reached from its parent. That means, for a tree with $n$ nodes, the function is called exactly $n$ times. In every function call exactly one swap operation is executed. Therefore, the worst-case number of copy operations is $\mathcal{O}(n)$.

\pathsort~has a significant advantage compared to the two \mergesort~variants. With \pathsort~it is possible to only reorder a subtree of the full tree. Due to \pathsort~using the structure of the tree as part of its execution structure, the algorithm can be started with an initial node which is not the root of the tree. The algorithm then reorders the subtree starting at that node.

Choosing the starting location to place the first node in \pathsort~has a significant influence on the resulting order. When starting \pathsort~with the root as the initial location $loc$ should be initialized as 0 to start the order at the beginning of the array. This might not always be the best option when only sorting a subtree, as this might result in destroying the order of another subtree.
Therefore, the location should be chosen as the current location of the starting node.

When the structure of the tree does not change between multiple reorder operations using \pathsort, reordering a subtree starting at the location of the starting node keeps the order of the rest of the tree intact. This is due to the way \pathsort~places nodes. After a tree was reordered using \pathsort, for any subtree all nodes of that subtree are in a continuous part of the array. That means no node outside the subtree is placed between nodes of the subtree. Therefore, this area of the array where the subtree is placed can be reordered arbitrarily, without moving any other node of the array.

If the structure of the tree did change, however, sorting a subtree might destroy the order of another subtree. The severity of this depends on the amount the structure of the subtree changed. This can be avoided by keeping track of whether the structure of the tree changed since the last reordering operation and potentially triggering a full tree reorder operation rather than a partial tree reorder operation.

\subsubsection{\Hpathsort~- Discussion}

While \pathsort~is designed to optimize performance in the \cmm, it might also be a viable strategy for a \gmm~application. One large benefit of the \pathsort~algorithm compared to the \mergesort~variants is its better worst-case number of copy operations, as it only increases linearly. In addition, only parts of the tree can be reordered.
While the resulting order is not optimal for a \gmm, it theoretically optimizes the most accessed paths, as most of the most accessed paths are placed in the more beneficial memory. Therefore, it can be expected to optimize the tree to an extent. The lower number of copy operations of \pathsort~might outweigh the shortcomings of the beneficial but not optimal order.

\subsection{Reducing the Number of Copy Operations}
\label{sec:reorder-improvements}

The explained reordering algorithms all utilized swap operations in their functions to move nodes inside the array to their correct location. As explained, copy operations and therefore swap operations are assumed to have the biggest impact on the overall performance. Reducing the number of copy operations is therefore a core objective when choosing the reordering algorithms.

However, the explained reordering algorithms do not use the minimal amount of copy operations to achieve their intended order. The minimal amount of copy operations for any new order copies every node to its intended location immediately, resulting in exactly as many copy operations as nodes need to be reordered. The true minimal amount of copy operations needs a few more copy operations and is discussed later, but for the sake of the argument it can be assumed that every node needs to be copied exactly once.

Even in the best case, where the proposed reordering algorithms place the nodes immediately in their intended location, this still results in more copy operations being executed. This is because the algorithms only use swap operations for reordering and swap operations inherently use three copy operations. Only in rare cases, where both nodes being swapped end up in their respective intended locations, this can result in the minimal number of copy operations. Otherwise, the number of copy operations is not minimal for the proposed algorithms.

Therefore, an additional way of reordering nodes is introduced. It does not implement any strategy of where to reorder the nodes to. This is still driven by the proposed strategies. Rather, it takes a discrete map of where the nodes are supposed to be placed and realizes that map in the minimal number of copy operations. To do this, it uses a variation of \cyclesort~\cite{DBLP:journals/cj/Haddon90}.

\subsubsection{\Hcyclesort}
\label{sec:reorder-improvements-principle}

\cyclesort~uses the observation, that in any finite discrete map of node locations, the node locations form cycles \cite{DBLP:journals/cj/Haddon90}. In a finite discrete map, each location maps to exactly one node location where this node should be placed. Furthermore, each node location is mapped to by exactly one location.

These maps from one location to another location form chains.
When following this chain of target locations the initial location will be reached at some point, thus forming a cycle.
Any map of $n$ node locations consists of any number of cycles between $1$ and $n$. The smallest cycle is one of a location mapping to itself. If all node locations map to themselves there are $n$ cycles. The largest cycle possible is one consisting of all locations resulting in exactly one cycle.

Any cycle of length $c = 1$ requires no copy operations, as the node is already in the correct location. Any cycle of length $c > 1$ needs exactly $c + 1$ copy operations to be realized. To prove this, it is observed that any node on a cycle of length $c > 1$ is not in its target location. If it were in its target location it would form a cycle of length one. Therefore, any cycle of length $c$ cannot be realized with less than $c$ copy operations. Otherwise, at least one node would not be copied into its target location. One additional copy operation is necessary, as the first node to be copied needs to be copied twice. First, it needs to be saved at another location outside the cycle, to make room for the node that should be placed in its location. In the end, it then needs to be copied into its target location.

This means, that for any map with $m$ cycles with a length larger than 1, the minimal number of copy operations needed to realize the map is $\sum_{i \in \{1,\dots, m\}} (c_{i} + 1)$, where $c_i$ is the length of cycle $i$.

The discrete map needs to encode where each node is supposed to be placed in the array. Fundamentally, the map is an array of values of the same length as the node array, where each value at each location encodes where a node is supposed to be placed. Two types of representations can be used to encode the new location. We call these two representations \repone~and \reptwo.
\repone~encodes the index of the node which should be placed in the location. \reptwo, in contrast, stores the index where the node at this location should be placed. The two representations are therefore the inverse of each other and one can be converted into the other.

\paragraph{Adapting \Hcyclesort}
\label{subsec:reorder-improvements-adapting}

In \cyclesort~developed by Haddon in \cite{DBLP:journals/cj/Haddon90} the algorithm is designed to sort the array according to its values. In one algorithm called ''special\_cycle\_sort'' it is assumed that the elements to sort are values from 1 to $n$ where $n$ is the length of the array to sort. Furthermore, no element is a duplicate, so every value from 1 to $n$ is present. The target location is therefore the value of the element itself and the used representation is \reptwo.

To work with the intended use case of this thesis, the algorithm is adapted to a different version called \mapreorder. Two fundamental changes are made to make \cyclesort~more suitable for the purposes of this thesis. Firstly, while the intended order directly comes from the element values in \cyclesort, for \mapreorder~the order is generated by the reordering strategies outlined.

Secondly, while \cyclesort~uses the \reptwo, the adapted version uses \repone~to further reduce the number of copy operations needed. Because \cyclesort~uses the \reptwo, it only knows where the current element should be placed, but not what should be placed at the location of the element. It therefore requires two copy operations per element. The adapted version using \repone, however, knows which element should be placed at the current location and can realize it in one copy operation (plus an additional at the beginning of the cycle).

\begin{algorithm}
    \Function{$\Fmapreorder(nodes, map)$}{
        $temp = \Ftempnode()$\;
        \ForEach{$i \textbf{ \emph{in} } \{0, \dots, \Fmaxindex()\}$}{
            \If{$map[i] \neq i$}{
                $temp.\Fcopy(nodes[i])$\;
                $cl = i; cs = map[i]$\;
                \While{$cs \neq i$}{ \label{line:map-reorder-while}
                    $nodes[cl].\Fcopy(nodes[cs])$\;
                    $map[cl] = cl; cl = cs; cs = map[cl]$\;
                } \label{line:map-reorder-while-end}
                $nodes[cl].\Fcopy(temp)$\;
                $map[cl] = cl$\;
            }
        }
    }
    \caption{\Hmapreorder} \label{alg:map-reorder}
\end{algorithm}

\Cref{alg:map-reorder} shows the \mapreorder~algorithm. It receives the nodes array and the map to be applied to the array as arguments.
It starts by reserving one node as temporary storage. This is used throughout the algorithm. It then iterates over the map, starting at the first index. If the map at the index is the same as the index, this means it is a cycle of length 1 and the node is already at the correct location. The node therefore does not need to be moved and is skipped.
When reaching an index where the two values do not match, a cycle of length $> 1$ has been found. This cycle has to be reordered.

In the first step, the node at the current location $i$ is moved to the temporary node location $temp$. Afterwards, a loop iterates over the cycle, moving the next node into its intended location. This is done until the original location is reached again. At this point the node stored in $temp$ can finally be moved into its intended location. During the loop, the map is updated whenever a node is moved to indicate that it is now in its intended location and does not need to be moved again.
This is repeated for every cycle until the end of the array is reached and all nodes have been placed.

\mapreorder~is designed to use the minimal amount of copy operations, to realize a given map as performant as possible.
The minimal number of copy operations required for applying any map with $m$ cycles is $\sum_{i \in \{1,\dots, m\}} (c_{i} + 1)$, where $c_i$ is the length of cycle $i$. For any cycle, \mapreorder~copies every node once to its intended location, except the start node, which is copied twice, once to a temporary location and then to its intended location. Therefore, it does use the minimal number of copy operations. The worst-case number of copy operations occurs when a map only consists of cycles of the length 2. That is because cycles of length 1 do not require any copy operations and the ratio of copy operations needed per length of cycle decreases the larger the cycle is. A map of length $n$ with only cycles of length 2 requires $\frac{n}{2}*3$ copy operations. That is because there exist $\frac{n}{2}$ cycles and every cycle requires $2 + 1 = 3$ copy operations.
Therefore, for a map of length $n$, the worst-case number of copy operations is $\mathcal{O}(\frac{n}{2}*3)$.

\subsubsection{\Hmapreorder~for the Reordering Strategies}

\mapreorder~is designed to realize any order in \repone. This means, as long as the reordering strategies outlined can be adapted to return their intended order in \repone, they can be combined with \mapreorder. For both \buffmergesort~and \shellmergesort~this can be done in a straightforward manner, by not sorting the array directly, but the order in \repone~and accessing the nodes for comparison through the map.

\pathsort~needs more care when being adapted. As the nodes are accessed through their pointer taken from the logical location in the tree, \repone~alone is not sufficient when creating the order. This is because the pointer can only be resolved to the index in the abstract array. But it needs to be known where the node is supposed to be placed in the current version of the map. This issue can be solved by using both \repone~and \reptwo~during map creation. This results in higher overhead compared to the two \mergesort~variants.
Furthermore, while \pathsort~can still be used to only reorder a subtree, a map for the whole array needs to be created, as it is not necessarily known in which parts of the array nodes have to be moved.

\section{Reordering During Tree Operations}
\label{cpt:during-reordering}

A tree can be reordered outside of tree operations without any additional changes to the explained reordering algorithms. However, the reordering strategies are designed to be used during tree operation. Even though the reordering algorithms are designed to be efficient, reordering still introduces an overhead. Reordering operations should therefore only be started if they are expected to improve performance significantly. To this end, two strategies are introduced to decide when to reorder.

As a first strategy \accthresh~is introduced.
\accthresh~checks how long it has been since a node was reordered during a reordering operation. Once the number of accesses since the last reorder operation exceeds a certain threshold, a reordering operation is triggered. It assumes that after enough tree operations, the \afm~has changed significantly enough to warrant the need for a reorder operation to be executed.
However, \accthresh~does not consider the changes in the distribution of the \afs.

In an attempt to consider the underlying changes in the \afm, when deciding whether a reorder operation is needed, \ratiothresh~is introduced. A reorder operation is only needed, if the overall distribution of the \afm~changed enough since the last reordering operation. This happens if the ratio of the \af~of children changes. For example, if the left child used to be accessed 80\% of the time, but now the right child is accessed 80\% of the time, a reorder operation could improve performance.

Importantly, when \pathsort~is used, rather than starting the reorder process from the node, the reordering process needs to be started from its parent. This is because the ratio amongst the children of the parent changed to a certain degree. Therefore, all siblings also need to be considered when reordering.

While this strategy considers the current distribution of the \afm, it is sensitive to small \af~values. The smaller the values in the \afm~are, the faster the ratios change. To prevent too many reorder operations due to small values in the \afm, this strategy additionally implements \accthresh~with a smaller threshold. Only if enough accesses were made to the node, the ratio is considered, and only if then the ratio changed enough, a reorder operation is started.

\section{Evaluation}
\label{cpt:evaluation}

To evaluate the proposed strategies, experiments are conducted to both test the performance of the reordering algorithms, as well as the impact of the achieved orders both as offline and online optimization. To this end, experiments are run on two types of systems. Firstly, a \fullmsp~(referred to as \msp~for brevity), which is equipped with SRAM and FRAM \cite{msptexas, msptexasdatasheet}. Of the two, SRAM is more performant both in terms of speed as well as energy efficiency \cite{msptexasdatasheet}. As such, the \msp~represents the \gmm. Secondly on a system equipped with an \fullepyc~(referred to as \epyc) with 256~GB of memory \cite{epyccpu}. The \epyc~has L1, L2 and L3 caches with sizes 8~MiB, 64~MiB and 512~MiB respectively. It therefore represents the \cmm.

In addition to the different systems, four different tree implementations were used. Firstly, \bstrees, which were picked as an example of a simple binary tree where the logical node locations do not change in relation to their physical locations \cite{DBLP:journals/jacm/Hibbard62}. Secondly, \avltrees, which are more complex binary trees, where nodes can be move around inside the structure of the tree \cite{adelson1962algorithm, avltreegithub}. Thirdly, \octrees, which are chosen because of their medium number of children, with up to 8 children per node \cite{DBLP:journals/cvgip/Meagher82, octreesource}. Lastly, \btrees, which have a lot of children per node \cite{DBLP:journals/acta/BayerM72, DBLP:journals/csur/Comer79}. All types of trees were adapted to use key-value pairs for more standardized experiments. The datasets were generated by using a random number generator to generate key value pairs that could be loaded into the tree as well as retrieved later. Furthermore, some key-value pairs were retrieved more often to simulate some key-value pairs being accessed more frequently than others.

\subsection{Reordering Algorithms}
\label{sec:evaluation-reorder}

As a first evaluation step, the different reordering algorithms are tested. Trees and their \afm~are generated and the nodes then shuffled. Afterwards, the different reordering algorithms are used to reorder the tree, while the time is measured. Overall, all types of trees showed similar results, therefore only the results for \avltrees~are discussed.

\buffmergesort~is depicted in \graphcolthree~with an unfilled square marking \nativebuffmergesort~(\SNbuffmergesort) and a filled square marking \mapbuffmergesort~(\SMbuffmergesort). Similarly, \shellmergesort~is shown in \graphcolfive, and marked with an unfilled circle for \nativeshellmergesort~(\SNshellmergesort) and a filled circle for \mapshellmergesort~(\SMshellmergesort). Lastly, \pathsort~is colored in \graphcolone, and marked with triangles with the unfilled being \nativepathsort~(\SNpathsort) and the filled being \mappathsort~(\SMpathsort).

\begin{figure}
    \centering
    \includegraphics{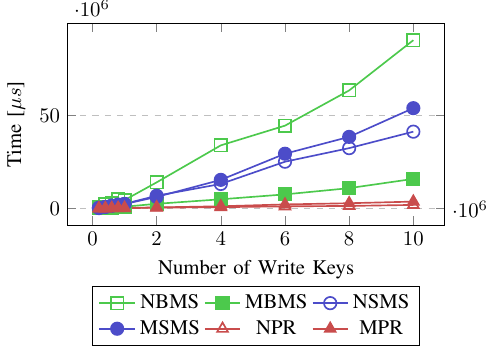}

    \caption{Reordering Strategies Runtime}
    \label{fig:eval-reorder-runtime-avl-shuffle}
\end{figure}

The runtime of the reordering algorithms can be seen in \Cref{fig:eval-reorder-runtime-avl-shuffle}. The graph shows the number of write keys, i.e. the size of the tree, on the x-axis and the total runtime of the reordering algorithms in $\mu s$ on the y-axis. \nativebuffmergesort~has the longest runtime. The second and third-longest runtimes are achieved by both variants of \shellmergesort. \nativeshellmergesort~has a shorter runtime than \mapshellmergesort. The fastest reordering algorithm of the \mergesort~strategy is \mapbuffmergesort. The fastest reordering algorithm overall is \pathsort, with both variants being significantly faster than any of the other reordering algorithms. Amongst the two variants, \nativepathsort~is faster than \mappathsort.

\begin{figure}
    \centering
    \includegraphics{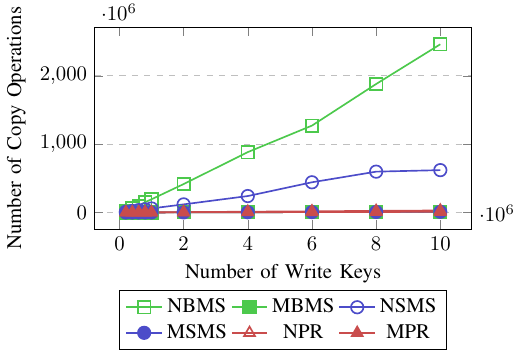}

    \caption{Reordering Strategies Copy Operations}
    \label{fig:eval-reorder-copy-avl-shuffle}
\end{figure}

To get a better understanding of the reason behind these runtimes, \Cref{fig:eval-reorder-copy-avl-shuffle} shows the same experiments reordering an \avltree, however with the number of copy operations executed plotted on the y-axis. As can be seen, both \nativebuffmergesort~and \nativeshellmergesort~need significantly more copy operations than the other four reordering algorithms. \nativebuffmergesort~needs almost 2.5 billion copy operations, while \nativeshellmergesort~needs over 500 million for a tree with 10 million nodes.

\begin{figure}
    \centering
    \includegraphics{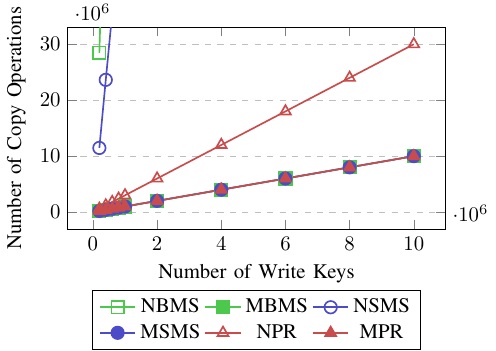}

    \caption{Reordering Strategies Copy Operations - Adapted Range}
    \label{fig:eval-reorder-copy-avl-shuffle-closer}
\end{figure}

As the number of copy operations of the other reordering algorithms cannot be properly seen in \Cref{fig:eval-reorder-copy-avl-shuffle}, \Cref{fig:eval-reorder-copy-avl-shuffle-closer} shows the same results, however with an adapted range of the y-axis. Here, it can be seen that all variants using \mapreorder~need almost the same number of copy operations. Furthermore, the number of copy operations rises linearly, being almost exactly the same as the number of write keys (i.e. the number of nodes in an \avltree).
Lastly, \nativepathsort~uses about three times as many copy operations as \mappathsort.

The results for the number of copy operations for the \mapreorder~variants show that \mapreorder~does use the minimal number of copy operations. As the tree is fully shuffled, almost all nodes need to be placed at another location inside the node array. This means the minimal number of needed copy operations is at least the number of nodes. Any additional copy operations are needed for the first node in a cycle. As the deviation from the number of nodes is small for all experiments (usually in the lower two digits), \mapreorder~appears to use the minimal number of copy operations.

\nativepathsort~needing three times the number of copy operations as the \mapreorder~variants also follows directly from the algorithm. Every node needs to be moved, however in \nativepathsort, this is done using a swap operation, which needs three copy operations. Therefore, for moving every node into its intended location, it needs three times the number of copy operations as \mapreorder.

The extremely high number of copy operations needed by \nativebuffmergesort~is due to its need to copy every node into the buffer for every recursion. This is likely the driving factor behind its higher runtime compared to the other reordering algorithms. However, \mapbuffmergesort~performs better than both \shellmergesort~variants. One factor is the low number of copy operations resulting from using \mapreorder. This is not the only factor, as \mapshellmergesort~also uses \mapreorder~but is slower than \mapbuffmergesort. This shows, that \buffmergesort~is more performant than \shellmergesort~when needing to sort elements which are less costly to copy, such as the integer values in the map.

\nativeshellmergesort~has a high number of copy operations. It needs fewer copy operations in the experiments compared to \nativebuffmergesort, even though it has a higher worst-case number of copy operations. However, as is explained, while the worst-case number of copy operations is higher, the average case is lower, as \nativebuffmergesort~always copies the array into the buffer. This effect is seen in the results of the experiments, as the experiments do not test the worst-case.

\subsection{Potential Performance Improvement}
\label{sec:evaluation-potential}

As a second evaluation step, the potential performance improvements are tested. The goal of these experiments is to show the potential impact that reordering the nodes can have on the overall performance. In the experiments, a tree is first filled with all write keys in the dataset. Afterwards, all read keys are read while generating the \afm. Then, two different kinds of experiments are run.
First, the entirety of the read keys are read again while measuring the time. This is done without any change in the order of the nodes. The other experiment first reorders the tree according to one of the two reordering strategies and then reads all read keys again while measuring the time. The two measured times can be compared to show the effect the different orders have on the performance.

\begin{figure}
    \centering
    \includegraphics{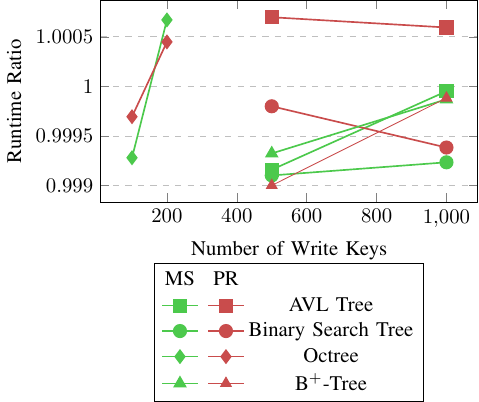}

    \caption{Potential Performance Improvement - Runtime (\msp)}
    \label{fig:eval-potential-ratio-msp-runtime}
\end{figure}

\Cref{fig:eval-potential-ratio-msp-runtime} shows the influence of the orders achieved by the two reordering strategies on the runtime performance for the different trees on the \msp. The runtime is improved in most experiments, the highest improvement is 0.1\%. For almost all trees, \mergesort~results in a larger performance improvement compared to \pathsort. Only when using \btrees, the opposite can be reported to be the case. Overall the improvement is very minor.

The experiments show that the reordering strategies are usually successful in improving performance of trees in the \gmm. \mergesort~improves the performance more than \pathsort. The intended goal of using \mergesort~to increase the accesses to beneficial types of memory is therefore achieved. However, the performance improvement is minor due to the small difference in the performance of SRAM and FRAM. As the two types of memory do not exhibit large differences in their performance, the potential performance is also equally low.

\begin{figure}
    \centering
    \includegraphics{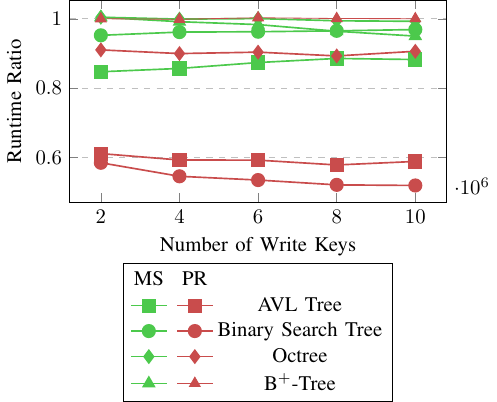}

    \caption{Potential Performance Improvement (\epyc)}
    \label{fig:eval-potential-ratio-avl-binary-oct}
\end{figure}

To analyze the impact the in the \cmm, \Cref{fig:eval-potential-ratio-avl-binary-oct} shows the runtime ratio of the experiments run on the \epyc. The x-axis shows the number of write keys and the y-axis shows the ratio of the runtimes. The results of trees being reordered with \mergesort~are colored in \graphcolthree, while the ones reordered using \pathsort~are colored in \graphcolone. \avltrees~are marked with squares, \bstrees~are marked with circles, \octrees~are marked with diamonds and \btrees~are marked with triangles.

Both reordering strategies decrease the needed runtime in most of the results.
\mergesort~increases performance of \avltrees~by up to 22\% in the best case. The performance of \bstrees~is improved by about 7.5\%, while the performance of \octrees~is increased by up to 3\%. The performance improvement diminishes, with larger trees for both \avltrees~and \bstrees.

\pathsort~consistently improves performance for all depicted types of trees. Especially \avltrees~and \bstrees~benefit from the order realized by \pathsort. \avltrees~are about 73\% faster when reordered with \pathsort, while \bstrees~are sped up by up to 94\%. \octrees~are not affected as strongly, but are still sped up by up by 16\%. Larger trees seem to slightly increase the performance improvement for \avltrees~and \bstrees, while they seem to not affect \octrees. \btrees~are affected the least, with the performance difference being within 1\%.

The results suggest the conclusion that thinner and deeper trees benefit most from the different order. Both \avltrees~and \bstrees~have two children per node and therefore grow very deep. \pathsort~increases performance on \bstrees~more than on \avltrees. However, in the experiments using \mergesort, the opposite is the case. The reason for this is likely that \bstrees~generally grow deeper than \avltrees. As \bstrees~are not balanced, they are capable of growing deeper with the same number of write keys compared to \avltrees. This results in \pathsort~having a larger impact, as the paths to place in memory are longer. \mergesort, in contrast, is not as beneficial for \bstrees, as the long paths result in a lower likelihood of the nodes accessed during one operation being placed closer together. Therefore, \mergesort~is more beneficial for \avltrees~compared to \bstrees.

Thinner and deeper trees favor \pathsort, as the paths can be longer and have higher frequencies. That is because, in a tree with more children per node, the frequency of every child is usually lower than in a tree with fewer children per node. For example, in a uniform \avltree, the ratio of each child node is 50\%, while in a uniform \octree~the ratio is 12.5\%. The lower fan out therefore results in frequently accessed paths usually having a higher frequency. Furthermore, the increased possible length helps \pathsort~further by resulting in longer sequential memory accesses.

As \pathsort~improves performance, this leads to the conclusion that the intended effect of improving cache locality and helping the CPU prefetch node contents is achieved. The results further show that \mergesort~can also improve performance in some instances. This is likely due to similar effects, as frequently accessed nodes are placed close together. However, it is not as effective as \pathsort.

\subsection{Reordering During Operation}
\label{sec:evaluation-during}

As a last evaluation step, the performance of the trees when reordering during operation is measured. 
Two types of performance data are gathered. First, a baseline execution, without any optimization or additional computation. That means the \afm~is not updated, it is not checked whether reordering should be done and no reordering operation is executed. It is therefore the standard execution of the tree, the only difference being that the nodes are also allocated to the same node array.
Secondly, the dataset is used, while optimizing the tree by reordering during operation. Both strategies for deciding when to reorder as explained in \Cref{cpt:during-reordering} are tested. As the potential performance improvement was minor on the \msp, no performance gain was measured during operation. On the \epyc, \mapbuffmergesort~also did not show any performance improvement. This is likely due to its lower potential performance improvement paired with a higher overhead when reordering. The following concentrates on results on the \epyc~using \nativepathsort.

\begin{figure}
    \centering
    \includegraphics{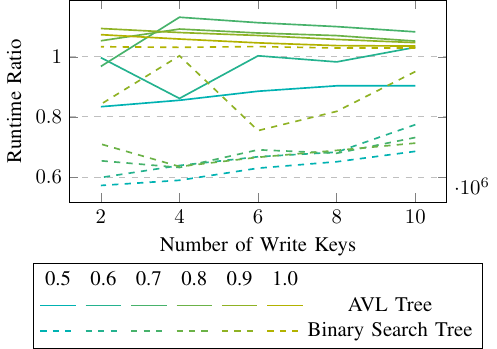}

    \caption{Reordering During Operation - \Hnativepathsort~- \bstree~\&~\avltree~- \Hratiothresh~(\epyc)}
    \label{fig:eval-during-bstree-avl-ratio}
\end{figure}

\begin{figure}
    \centering
    \includegraphics{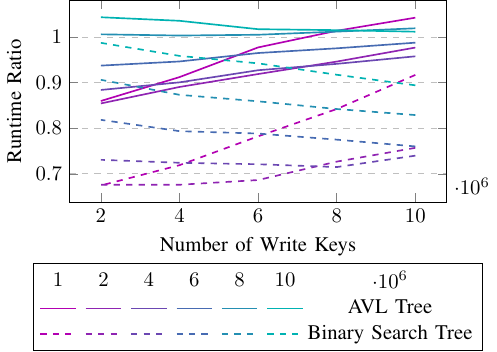}

    \caption{Reordering During Operation - \Hnativepathsort~- \bstree~\&~\avltree~- \Haccthresh~(\epyc)}
    \label{fig:eval-during-bstree-avl-notratio}
\end{figure}

\begin{figure}
    \centering
    \includegraphics{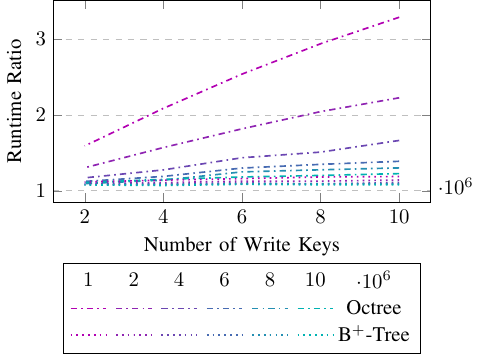}

    \caption{Reordering During Operation - \Hnativepathsort~- \octree~\&~\btree~- \Haccthresh~(\epyc)}
    \label{fig:eval-during-oct-btree-notratio}
\end{figure}

The first results to examine are using \ratiothresh~on \avltrees~and \bstrees. \Cref{fig:eval-during-bstree-avl-ratio} shows the results, with the x-axis showing the number of write keys and the y-axis showing the runtime ratio. Furthermore, a full line shows the results for \avltrees~and a dashed line for \bstrees. Lastly, the color shows the threshold used for \ratiothresh~with a more \graphcoltwo~line showing a lower value and a more \graphcolfour~showing a higher value.

\accthresh~manages to improve performance in many configurations for both \avltrees~and \bstrees. Especially \bstrees~see a significant performance improvement of up to 75\%. It is especially apparent, that lower thresholds yield greater performance improvements than lower thresholds.

This leads to the conclusion that frequent reordering improves performance significantly in binary trees. The overhead introduced by more frequent reordering is outweighed by the performance gain. \bstrees~show higher performance improvements, likely because they are less sensitive to changes in the ratio of the \af~of children. That is because the logical structure of an \bstree~never changes, while \avltrees~have balancing operations, which can change the ratios significantly.

\accthresh~shows similar results. Everything in \Cref{fig:eval-during-bstree-avl-notratio} is marked the same way as in \Cref{fig:eval-during-bstree-avl-ratio} with the only exception being that a \graphcolfour~shows results for a lower threshold and \graphcolsix~for a higher. Overall, the results are similar with \bstrees~showing a higher performance improvement. Overall, \bstrees~show a lower performance improvement compared to \ratiothresh. \avltrees~show the inverse result, with \accthresh~showing higher performance improvements. The reason for this is that \accthresh~is less sensitive to the structural changes in \avltrees. Overall it can be reported that reordering during operation can improve performance for binary trees.

Trees with higher number of children as \octrees~and \btrees~show different results. \Cref{fig:eval-during-oct-btree-notratio} shows the results for \octrees~and \btrees. The results show that no performance improvement was achieved. This is like due to the lower potential performance improvement of shallow and wider trees. That means, the performance gain is not large enough to offset the overhead of reodering.

\section{Conclusion}
\label{cpt:conclusion}

This paper examines ways to use reordering of tree nodes inside memory to benefit performance. For this,
different strategies of reordering nodes inside memory are introduced. Firstly, different variants of \mergesort, which aim to increase the accesses to more beneficial types of memory while decreasing the accesses to less beneficial types of memory in the \gmm. Secondly, \pathsort, a reordering strategy which places frequently accessed paths successively inside memory to increase spatial locality in the \cmm. Efficient algorithms realizing these strategies are developed. All algorithms are also expanded by using \mapreorder~to further reduce the number of copy operations.

These strategies are then evaluated using a set of experiments. The experiments show that \mapbuffmergesort~performs best of all the \mergesort~variants. Furthermore, the overall most performant reordering algorithm is \nativepathsort. The performance of most types of trees can be improved by means of a new order of the nodes inside memory. In the \gmm~only a minor improvement in performance is achieved, due to the difference in the different types of memory being small to begin with. The experiments in the \cmm~show promising performance results. Especially deep and thin trees, such as \bstrees~and \avltrees, benefit most from a new order, reaching a performance improvement of up to 95\%. Wider trees, such as \octrees~or \btrees, show less performance gain.

Lastly, the experiments further show that for certain configurations, \nativepathsort~can be used to improve the performance of trees during operation by reordering the tree according to certain strategies. \octrees~and \btrees~cannot be optimized during operation with the strategies explained in this paper, however, \avltrees~and \bstrees~show significant performance improvement when reordered during operation. Choosing the correct strategy for deciding when to reorder as well as a fitting threshold is key to achieving the highest possible performance improvement. Overall, the largest performance improvement when reordering during operation is a speedup of 75\% using \bstrees.

To conclude, this paper shows the potential performance benefits of reordering nodes inside memory. It further shows that online optimization of tree-based data structures using the reordering strategies can be successful. The overall goal of this paper of optimizing tree-based data structures can therefore be reported to have been achieved.

\section*{Acknowledgement}
This paper has been supported by
Deutsche Forschungsgemeinshaft (DFG), as part of the project
Memory Diplomat (502384507) within SPP 2377:  Disruptive Main-Memory Technologies.

\bibliographystyle{IEEEtran}
\bibliography{IEEEabrv,literatur}

\end{document}